\def\year{2018}\relax
\documentclass[letterpaper]{article} 
\usepackage{aaai18}  
\usepackage{times}  
\usepackage{helvet}  
\usepackage{courier}  
\usepackage{url}  
\usepackage{graphicx}  

\usepackage{mathptmx}
\usepackage{amssymb}
\usepackage{amsmath}
\usepackage{subfigure}
\usepackage{hyphenat}
\usepackage{color}
\usepackage{textcomp}
\usepackage{enumitem}
\usepackage{verbatim}
\usepackage{algorithm}
\usepackage{algorithmic}
\usepackage{enumitem}
\usepackage{pifont}
\usepackage{xspace}
\usepackage{tabu} 

\usepackage[show]{chato-notes} 

\usepackage{cmm-greek}

\newcommand{\hide}[1]{}
\newcommand{\xhdr}[1]{\vspace{1.7mm}\noindent{{\bf #1.}}}

\newcommand{\ie}{\textit{i.e.}\xspace}
\newcommand{\eg}{\textit{e.g.}\xspace}
\newcommand{\cf}{\textit{cf.}\xspace}

\newcommand{\vs}{\textit{vs.}\xspace}
\newcommand{\etc}{\textit{etc.}\xspace}

\newcommand{\name}{Quootstrap\xspace}

\newcommand{\githubrepo}{\url{https://github.com/epfl-dlab/quootstrap}}

\definecolor{darkred}{RGB}{190, 53, 53}
\definecolor{darkgreen}{RGB}{53, 160, 53}
\definecolor{darkblue}{RGB}{53, 53, 190}
\newcommand{\Qcol}[1]{\textcolor{darkred}{#1}}
\newcommand{\Spcol}[1]{\textcolor{darkgreen}{#1}}
\newcommand{\wccol}[1]{\textcolor{darkblue}{#1}}
\newcommand{\wildcard}{\wccol{$*$}\xspace}
\newcommand{\Q}{\textit{\Qcol{Q}}\xspace}
\newcommand{\Sp}{\textit{\Spcol{S}}\xspace}
\newcommand{\quot}[1]{{\fontfamily{phv}\selectfont #1}}
\newcommand{\pattern}[1]{\quot{[#1]}}
\newcommand{\ellipsis}{[...]\xspace}

\newcommand{\patprec}{\pi}
\newcommand{\qsp}{quotation--speaker pair\xspace}

\newcommand{\Secref}[1]{Sec.~\ref{#1}}
\newcommand{\Eqnref}[1]{Eq.~\ref{#1}}

\newcommand{\Tabref}[1]{Table~\ref{#1}}
\newcommand{\Figref}[1]{Fig.~\ref{#1}}

\newcommand{\Appref}[1]{Appendix~\ref{#1}}

\newcommand{\specialcell}[2][c]{\begin{tabu}[#1]{@{}c@{}}#2\end{tabu}} 

\begin{document}

\title{Quootstrap: Scalable Unsupervised Extraction of Quotation--Speaker Pairs from\\Large News Corpora via Bootstrapping}

\author{
Dario Pavllo\\
EPFL\\
dario.pavllo@epf\/l.ch
\And
Tiziano Piccardi\\
EPFL\\
tiziano.piccardi@epf\/l.ch
\And
Robert West\\
EPFL\\
robert.west@epf\/l.ch
}

\maketitle

\begin{abstract}
We propose \name, a method for extracting quotations, as well as the names of the speakers who uttered them, from large news corpora.
Whereas prior work has addressed this problem primarily with supervised machine learning, our approach follows a fully unsupervised bootstrapping paradigm.
It leverages the redundancy present in large news corpora, more precisely, the fact that the same quotation often appears across multiple news articles in slightly different contexts.
Starting from a few seed patterns, such as [``$Q$'', said $S$.], our method extracts a set of quotation--speaker pairs $(Q,S)$, which are in turn used for discovering new patterns expressing the same quotations; the process is then repeated with the larger pattern set.
Our algorithm is highly scalable, which we demonstrate by running it on the large ICWSM 2011 Spinn3r corpus.
Validating our results against a crowdsourced ground truth, we obtain 90\% precision at 40\% recall using a single seed pattern, with significantly higher recall values for more frequently reported (and thus likely more interesting) quotations.
Finally, we showcase the usefulness of our algorithm's output for computational social science by analyzing the sentiment expressed in our extracted quotations.
\end{abstract}

\section{Introduction}
\label{sec:intro}
Online news constitutes a primary source of information for the masses.
In addition to simply documenting the course of global and local events, news sources also serve as an archive of who said what, by means of the quotations embedded into news articles.

Quotations are interesting and important because they directly capture the opinions of those who have uttered them.
Quotations have therefore played an eminent role in computational analyses of the media landscape and news cycle.
For instance, \citeauthor{niculae2015quotus} (\citeyear{niculae2015quotus}) analyze how President Obama's State of the Union address was quoted in news articles;
\citeauthor{gentzkow2016measuring} (\citeyear{gentzkow2016measuring}) measure polarization in the U.S.\ political landscape by analyzing quotations uttered by members of Congress;
and the Memetracker \cite{leskovec2009meme} and NIFTY \cite{suen2013nifty} projects aim to quantify the temporal dynamics of the news cycle by analyzing how quotations propagate from outlet to outlet.

The first two of the aforementioned projects consider narrow contexts (one speech by Obama and speeches given in Congress, respectively) and work with datasets where quotations have been linked to their respective speakers.
The latter two projects, on the contrary, consider broad contexts (millions of quotations appearing on thousands of websites), but their datasets consist of quotations only, without any information on who uttered them.

What would be tremendously useful, but is currently not available, is a corpus combining the advantages of all these settings: a Web-scale corpus of quotations attributed to speakers.
Use cases of such a dataset abound:
we could use it in temporal analyses to understand how the language of politics has changed over the years (\eg, is it true that the language of politics has become more vitriolic with time?);
we could use it to construct a mention graph of who mentions whom in order to understand the social structure of political discourse;
combining the mention graph with sentiment analysis could be useful in detecting coalitions;
and finally, a Web-scale dataset of \qsp{}s could also aid tasks such as fact checking, bias detection, \etc

The task of extracting quotations and attributing them to speakers is complicated by the wide range of ways in which a quotation can be expressed.
The task has been addressed according to two broad paradigms:
approaches based on \textit{supervised machine learning} train a classifier on sets of manually annotated \qsp{}s,
whereas approaches based on \textit{unsupervised pattern matching} extract pairs using regular expressions or similar rule\hyp based techniques, where patterns may be handcrafted or learned from data.

Each approach comes with its own advantages and disadvantages. Supervised learning techniques tend to generalize better and cover more cases but have the disadvantage of requiring a manually labeled training set, which is typically laborious to obtain.
Moreover, models derived via machine learning are often hard to interpret, and their scope is restricted to the single language on which they were trained.
Approaches based on handcrafted patterns, on the contrary, often have high precision in practice, but their recall tends to be low, unless a large number of patterns are used. Similarly to data labeling in supervised machine learning approaches, hand\hyp coding regular expressions is a tedious process.

\name, the quotation extraction and attribution algorithm presented in this paper, falls into the category of pattern\hyp matching--based approaches, but unlike other approaches (\eg, \citeauthor{pouliquen2007automatic} \citeyear{pouliquen2007automatic}) it does not require fastidious pattern engineering.
All we need is a very small number of initial \textit{seed patterns}---in fact, as we will see, a single seed pattern suffices.

\begin{figure}
	\centering{}
	\includegraphics[width=\linewidth]{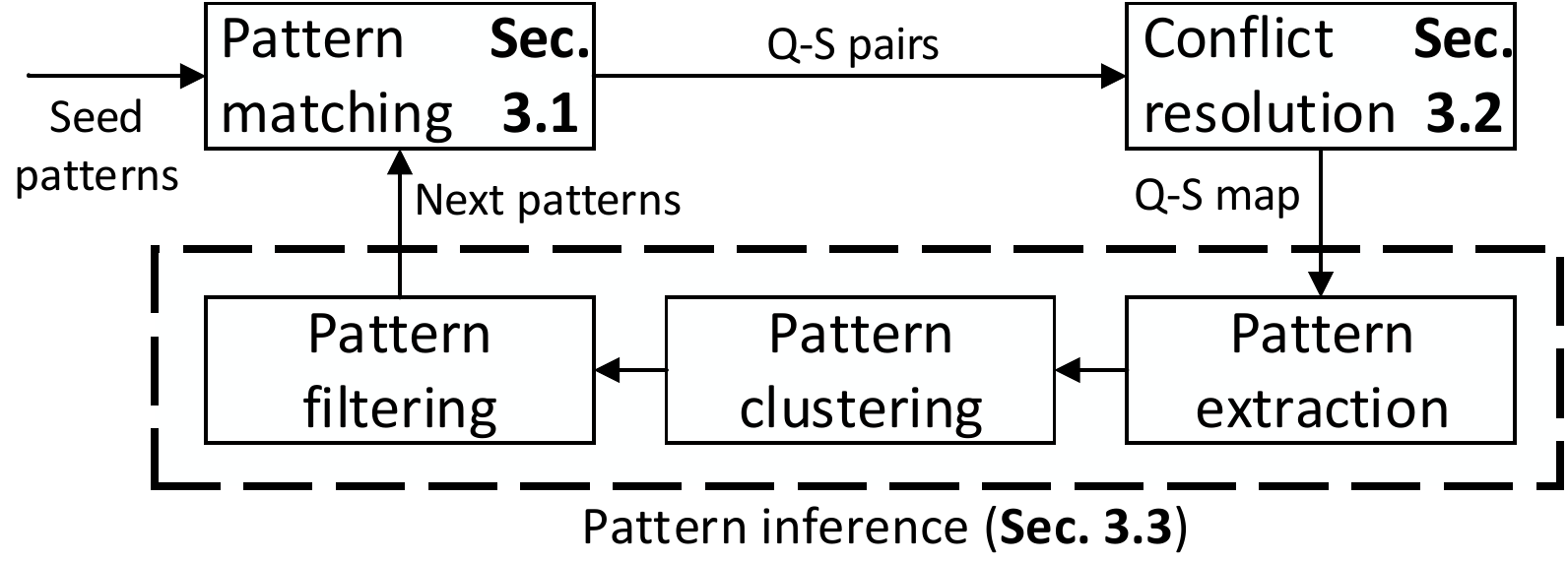}
	\caption{The \name loop, our core contribution.
    Individual components are annotated with the sections of the paper in which they are discussed.
    ``Q-S'' stands for ``quotation--speaker''.
    }
	\label{fig:pipeline}
\end{figure}

\xhdr{\name{} in a nutshell}
Our algorithm is a conceptually simple variant of a general paradigm called \emph{bootstrapping} \cite{Hearst1992}.
It starts with a set of handcrafted, high\hyp precision (but likely low\hyp recall) \emph{seed patterns} such as \pattern{\Q, said \Sp.} or \pattern{\Q, \Sp said} (where \quot{\Q} stands for a quotation, and \quot{\Sp} for a speaker).
The seed patterns are used for extracting quotation--speaker pairs from the dataset.
Afterwards, the algorithm finds all occurrences of these pairs in the corpus and uses them for inferring new patterns.
The new patterns are then used for extracting more pairs, and this alternating process is repeated iteratively.

\xhdr{Example}
Consider a toy example, where the (tiny) corpus consists of five short documents:
\begin{enumerate}
\item \quot{\ellipsis \Qcol{``I love \name''}, said \Spcol{Queequeg}. \ellipsis}
\item \quot{\ellipsis \Qcol{``Oops''}, said Mr.\ \Spcol{Melville}. \ellipsis}
\item \quot{\ellipsis \Qcol{``I love \name''}, said Mr.\ \Spcol{Queequeg}. \ellipsis}
\item \quot{\ellipsis \Spcol{Queequeg}, who appears in \Qcol{``Moby Dick''}, \ellipsis}
\item \quot{\ellipsis \Qcol{``I love \name''}, said the queasy \Spcol{Queequeg}.}
\end{enumerate}

Also assume we start with a single seed pattern, \pattern{\Q, said \Sp.}.
In this scenario, the algorithm would match only document~1, producing the pair $(Q=\text{\quot{\Qcol{``I love \name''}}}, S=\text{\quot{\Spcol{Queequeg}}})$. Then, it would find all contexts in which this pair appears (except for the ones that have already returned a result), \ie,
\begin{enumerate}
\item[3.] \quot{\Qcol{``I love \name''}, said Mr.\ \Spcol{Queequeg}.}
\item[5.] \quot{\Qcol{``I love \name''}, said the queasy \Spcol{Queequeg}.}
\end{enumerate}
Finally, the quotation and the speaker would be replaced with their respective placeholders so as to obtain the new patterns \pattern{\Q, said Mr.\ \Sp.} and \pattern{\Q, said the queasy \Sp.}.
Following this, the pair $(Q=\text{\quot{\Qcol{``Oops''}}}, S=\text{\quot{\Spcol{Melville}}})$ would be discovered in document~2 in the next iteration, and so forth.

The main advantages of \name consist in its scalability and in the fact that it is unsupervised and does therefore not require any labeled training data.
The latter aspect makes our method language\hyp independent: whenever we want to run it on a corpus in a new language, all we need is a very small set of handcrafted seed patterns.

A potential downside is the inability of resolving coreferences (\eg, \quot{he said}, rather than \quot{\Spcol{Queequeg} said}). We will show, however, that this is less of a problem than it may seem at first, as a large fraction of quotations is replicated across multiple articles, such that we can often find at least one explicit speaker attribution amenable to \name's extraction patterns (\Secref{sec:Impact of not resolving coreferences}).

The remainder of this paper is structured as follows.
We start by discussing related work (\Secref{sec:relwork}).
We then explain the \name algorithm in \Secref{sec:method} (technical details can be found in the appendix), describe the news corpus we use for building and evaluating our extraction patterns in \Secref{sec:data},
and discuss the results we obtain in \Secref{sec:results}.
In \Secref{sec:apps}, we showcase the utility of the extracted quotation corpus in an analysis of speaker sentiment, and we conclude the paper with a discussion in \Secref{sec:discussion}.

We make all of our code and data publicly available at \githubrepo.

\section{Related work}
\label{sec:relwork}
As mentioned, one of the two principal approaches to quotation attribution is \textit{supervised machine learning} trained on a hand-annotated ground truth.
\citeauthor{muzny2017two} (\citeyear{muzny2017two}) propose a sieve-based approach specifically designed for the literary domain,
while \citeauthor{o2012sequence} (\citeyear{o2012sequence}) develop a model based on sequence labeling that can handle both literary and news text.
Further supervised learning approaches are due to \citeauthor{glass2007naive} (\citeyear{glass2007naive}), \citeauthor{elson2010automatic} (\citeyear{elson2010automatic}),
and \citeauthor{almeida2014joint} (\citeyear{almeida2014joint}).
The latter approach is particularly interesting in that it combines quotation attribution and coreference resolution in a joint model.

The other principal approach---\textit{pattern-based quotation attribution}---has been studied by \citeauthor{pouliquen2007automatic} (\citeyear{pouliquen2007automatic}), who develop a system capable of extracting \qsp{}s from news articles.
Their implementation uses a set of handcrafted patterns in 11 languages.

To the best of our knowledge, no previous approach has exploited the redundancy of news text, where many sources tend to publish the same quotations in slightly different contexts.
Leveraging this fact allows our pattern\hyp matching--based method to achieve good results even with a single seed pattern, by adhering to the \textit{bootstrap} paradigm first introduced by \citeauthor{Hearst1992} (\citeyear{Hearst1992}) as a strategy to discover hyponym relations from a large text corpus.
It is important to note that Hearst's contribution was a meta algorithm, rather than a concrete implementation:
she left the most challenging step---inferring a small set of patterns from a set of previously matched contexts---to be performed manually by a human in the loop.

Several years later, Brin (\citeyear{brin1998extracting}) proposed an automated implementation of this step, in his system for extracting relations (in particular, book--author pairs) from the Web.
Shortly after, \citeauthor{agichtein2000snowball} (\citeyear{agichtein2000snowball}) refined this technique with their \emph{Snowball} system, which included a method for filtering discovered patterns and thus ensuring high\hyp precision matches throughout all iterations of the bootstrap loop.

\name is inspired by Snowball, but differs from it in several aspects.
For instance, our goal is to extract \qsp{}s, a task that poses additional challenges, compared to the simpler relation extraction tasks handled by Snowball, such as the coreference issue described above, as well as the fact that the same quotation may appear in slightly different or abridged forms throughout the corpus.
Finally, we developed a more sophisticated pattern inference procedure: while that of Snowball reasons about text as a bag of words, we maintain word order and develop efficient algorithms for this setting.

\section{Method}
\label{sec:method}

We have summarized the overall idea of our bootstrapping approach, and illustrated it with an example, in the introduction (\Secref{sec:intro}).
It essentially works by alternating two dual steps:
(1)~using known patterns to extract new \qsp{}s from the corpus, and
(2)~using known \qsp{}s to extract new patterns.

\xhdr{Challenges}
Based on the example from the introduction, we point out three issues that may arise:
\begin{enumerate}
\item Overly specific patterns, \eg, \pattern{\Q, said the queasy \Sp.}: although likely to identify the correct speaker, such patterns will match only very rarely and therefore represent a computational burden without much payoff (high precision, low recall).
\item Overly general patterns, \eg, \pattern{\Sp, who appears in \Q{}}, which (with a different set of seed patterns) might be extracted from document~4 in \Secref{sec:intro}: this pattern may appear with any \textit{Moby Dick} character, rather than just Queequeg (\eg, \quot{\Spcol{Captain Ahab}, who appears in \Qcol{``Moby Dick''}}), and is therefore not well suited for extracting correct speakers, leading to low precision (note that \quot{\Qcol{``Moby Dick''}} is not even a proper quotation, but our method is agnostic to that).
\item Conflicting quotation--speaker pairs: the same quotation may be attributed to several speakers; \eg, many people might say, \quot{\Qcol{``I love \name''}}.
\end{enumerate}

Addressing these issues is central for the success of our approach.
Otherwise, errors that arise in one iteration of the algorithm will be propagated to all subsequent iterations.
As a design principle, we aim to optimize for precision first (that is, the correctness of the extracted pairs), and then increase recall (the coverage of the patterns) via a union of many high\hyp precision patterns.

\xhdr{The \name{} loop}
The following description (visualized in \Figref{fig:pipeline}) delineates the sequence of steps involved in \name{};
it also mentions how we overcome the three aforementioned challenges.
Then, we give a more detailed description of each step in separate subsections.

\begin{enumerate}[start=0]
\item \textbf{Preprocessing:} Tokenize \cite{manning-EtAl:2014:P14-5}; detect speaker names and quotations; group variants of the same quotation (\cf\ \Appref{sec:Implementation details}); define a set of seed patterns and use it as the pattern set $\mathcal{P}$ in the first iteration.
\item \textbf{Pattern matching (\Secref{sec:Pattern matching}):} The patterns from the set $\mathcal{P}$ are matched against the corpus, and all the matching quotation--speaker pairs are returned.
\item \textbf{Conflict resolution (\Secref{sec:Conflict resolution}):} Pattern matching may extract the same quotation multiple times, and it may be attributed to different speakers. This step resolves such conflicts, attributing each quotation to at most one speaker and thereby addressing challenge~3 from above.
\item \textbf{Pattern inference (\Secref{sec:Pattern inference}):} All occurrences of the resulting \qsp{}s $(Q,S)$ are found in the corpus and are used for extracting new patterns. This step comprehends three substeps:
\textbf{Pattern extraction} locates all $(Q,S)$ pairs in the corpus and generates a candidate pattern from each occurrence.
\textbf{Pattern clustering} converts these patterns to a set of more general patterns, which increases recall and thereby addresses challenge~1 from above.
\textbf{Pattern filtering} discards overly general patterns, which increases precision and thus addresses challenge~2 from above.
\item Go to step~1, using the newly discovered patterns from step~3 as the pattern set $\mathcal{P}$.
\end{enumerate}

The following subsections discuss the above steps 1--3 in more detail.

\subsection{Pattern matching}
\label{sec:Pattern matching}

This step matches each pattern $P \in \mathcal{P}$ against the corpus, thus extracting a number of quotation--speaker pairs $(Q,S)$.

\xhdr{Pattern format}
Our patterns may be regarded as a restricted version of regular expressions. Our matching units are \emph{word tokens}, rather than characters.
Patterns must contain exactly one \emph{speaker placeholder} \quot{\Sp}, and exactly one \emph{quotation placeholder} \quot{\Q}.
Both \quot{\Sp} and \quot{\Q} can match one or more contiguous word tokens.
A pattern may contain \emph{wildcards} \quot{\wildcard}, each of which matches exactly one token.
A pattern must not start or end with \quot{\wildcard}, since this would be equivalent to the same pattern without boundary wildcards, nor with \quot{\Sp},
to avoid issues with multiple aliases of the same speaker name.

When a pattern is matched, the word sequences matched by the quotation and speaker placeholders become $Q$ and $S$, respectively, of the resulting pair $(Q,S)$.

\xhdr{Pattern precision}
For resolving conflicts and eliminating overly general patterns, we will rely on a notion of \textit{pattern precision,} which we introduce next.
Consider a quotation--speaker pair $(Q,S)$ extracted by a pattern $P$ in iteration $i$ of our algorithm.
Quotation $Q$ might already have been discovered in earlier iterations, by patterns other than $P$.
If this is the case, and if the speaker associated with quotation $Q$
before iteration $i$ is equal to (different from) $S$,%
\footnote{Recall that conflict resolution (\Secref{sec:Conflict resolution}) ensures that each quotation $Q$ is associated with at most one speaker $S$.}
then we refer to $(Q,S)$ as a \emph{positive (negative) match} of pattern $P$.
If $Q$ has not been found in previous iterations, then we refer to $(Q,S)$ as a \emph{neutral match} of $P$.
Now, the \emph{pattern precision} $\patprec(P)$ of a given pattern $P$ is defined as
\begin{eqnarray}
\label{eqn:patprec}
\patprec(P) &:=& \frac{n_+(P)}{n_+(P) + n_-(P)},
\end{eqnarray}
where $n_+(P)$ is the number of positive matches of pattern $P$, and $n_-(P)$ is the number of negative matches. This metric will prove useful in \Secref{sec:Conflict resolution} and \Secref{sec:Pattern inference}.
When $P$ is a seed pattern (\ie, $i=1$), the above definitions do not apply, so we let the precision of seed patterns be 1 by definition.


\subsection{Conflict resolution}
\label{sec:Conflict resolution}

It sometimes happens that different speakers are extracted for the same quotation $Q$ from different text passages. This may happen for a number of reasons:
\begin{enumerate}
\item $Q$ is truly attributed to different speakers by different texts, maybe because the quotation is common, or maybe because a source reports wrong information.
\item An imprecise pattern has led to an extraction error.
\end{enumerate}

To ensure high-quality results, we err on the conservative side and assume the latter, which is why we attempt to resolve such conflicts. In doing so, we exploit our notion of pattern precision (\Eqnref{eqn:patprec}). Consider a \qsp $(Q,S)$ extracted by one or more patterns; let $\mathcal{P}_{QS}$ be the set of patterns that have extracted $(Q,S)$. Inspired by \emph{Snowball} \cite{agichtein2000snowball}, we quantify our \emph{confidence} $\kappa(S;Q)$ in $S$ being the speaker who uttered $Q$ via the probability that at least one of the patterns that extracted $(Q,S)$ was correct in doing so,
\begin{eqnarray}
\kappa(S;Q) &:=& 1 - \prod_{P \in \mathcal{P}_{QS}}{(1 - \patprec(P))},
\end{eqnarray}
and assign quotation $Q$ to the speaker $S$ with the highest confidence $\kappa(S;Q)$. In the case of ties, as can easily happen in the first iteration (where all patterns have a precision of 1), we attribute the quotation to the most frequent speaker.

\subsection{Pattern inference}
\label{sec:Pattern inference}
Pattern inference represents the most complex step of our pipeline and is itself composed of three substeps (pattern extraction, clustering, and filtering), which we discuss in detail next.

The input to this step is the set $\mathcal{Q}$ of $(Q, S)$ pairs with all conflicts resolved, and the output is the set of new patterns that will be used in the next iteration.

\xhdr{Pattern extraction}
First, for each pair $(Q,S) \in \mathcal{Q}$, we extract the contexts where speaker $S$ appears near quotation $Q$, and we generate a \emph{minimal} valid pattern, \ie, the smallest pattern encompassing both the quotation and the speaker and meeting the format criteria of \Secref{sec:Pattern matching}.
For instance, the pair $(Q=\text{\quot{\Qcol{``I love \name''}}}, S=\text{\quot{\Spcol{Queequeg}}})$ might retrieve the context
\begin{quote}
\quot{According to government sources, Mr.\ \Spcol{Queequeg} enthusiastically proclaimed: \Qcol{``I love \name''}.}
\end{quote}
This match is then transformed into the pattern \pattern{Mr.\ \Sp enthusiastically proclaimed: \Q{}}.
The set of all patterns extracted for all pairs in $\mathcal{Q}$ is used as the pattern set $\mathcal{P}$ in the next substep.

\xhdr{Pattern clustering}
In the pattern set $\mathcal{P}$ generated by the above step, some patterns are immediately useful, \eg, \pattern{Mr.\ \Sp proclaimed: \Q{}}, while others are overly specific; \eg, \pattern{Mr.\ \Sp enthusiastically proclaimed: \Q{}} seems inefficient, as we might imagine any of a large number of adverbs taking the place of \quot{enthusiastically} (\eg, \quot{sadly}, \quot{loudly}, \quot{quickly}).
Hence, we aim to transform such patterns into a more general form, by preserving tokens that appear frequently and replacing with wildcards \quot{\wildcard} the infrequent ones.

Concretely, we start by constructing a so-called \emph{directed acyclic word graph} (DAWG) \cite{daciuk2000incremental} from the pattern set $\mathcal{P}$.
A DAWG is a deterministic acyclic finite state automaton that can be used to store a set of strings in compressed form.
The original strings can be reconstructed without loss of information by traversing the DAWG in a depth-first search from the root.
DAWGs can be constructed efficiently in an incremental fashion \cite{daciuk2000incremental}.
When building a DAWG for the patterns in $\mathcal{P}$, we also count how frequently each prefix (corresponding to a path starting in the root) occurs in the corpus, which allows us to recognize infrequent patterns and generalize them via wildcards.

\begin{figure}[t]
	\centering{}
	\includegraphics[width=\linewidth]{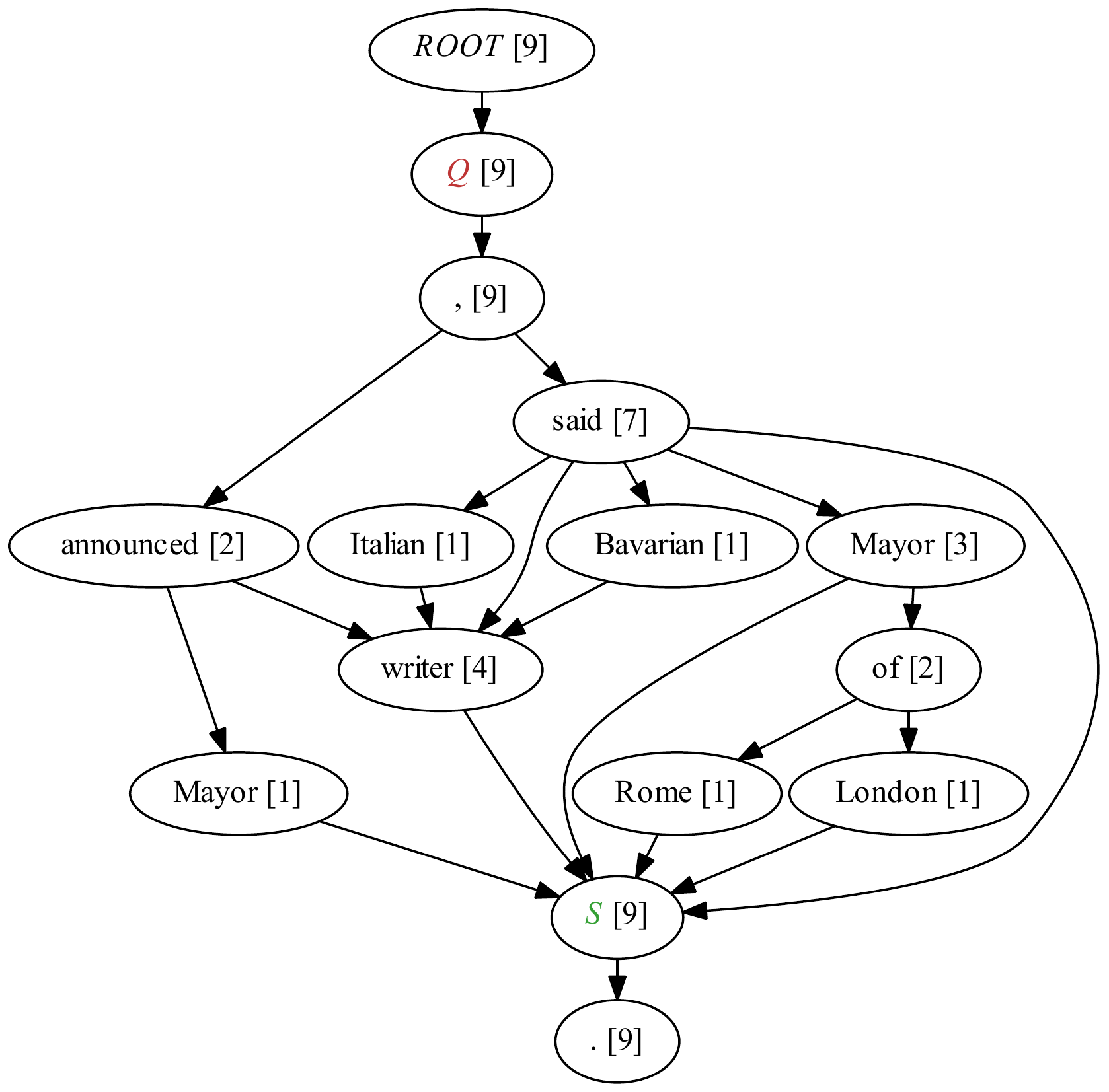}
	\caption{Directed acyclic word graph (DAWG) associated with the set of candidate patterns listed in \Secref{sec:Pattern inference}.
    Occurrence counts in square brackets.}
	\label{fig:wordGraph}
\end{figure}

As an example, \Figref{fig:wordGraph} depicts the DAWG corresponding to the following pattern set $\mathcal{P}$:
\begin{enumerate}
\item \pattern{\Q, said \Sp.}
\item \pattern{\Q, said writer \Sp.}
\item \pattern{\Q, said Italian writer \Sp.}
\item \pattern{\Q, said Bavarian writer \Sp.}
\item \pattern{\Q, announced writer \Sp.}
\item \pattern{\Q, announced Mayor \Sp.}
\item \pattern{\Q, said Mayor \Sp.}
\item \pattern{\Q, said Mayor of Rome \Sp.}
\item \pattern{\Q, said Mayor of London \Sp.}
\end{enumerate}

Given a threshold $n_{\min}$, our pattern clustering algorithm now replaces all DAWG vertices whose count is less than $n_{\min}$ 
with a wildcard \quot{\wildcard}.
Then, we extract all patterns from the DAWG via a depth-first search, trim leading and trailing wildcards (to be consistent with the pattern format constraints of \Secref{sec:Pattern matching}), and discard patterns with more than $m$ consecutive wildcards (we use $m=5$).
In the above example, a threshold of $n_{\min}=2$ results in the following patterns:
\begin{enumerate}
\item[1.] \pattern{\Q, said \Sp.}
\item[2.] \pattern{\Q, said writer \Sp.}
\item[3--4.] \pattern{\Q, said \wildcard writer \Sp.}
\item[5--6.] \pattern{\Q, announced \wildcard \Sp.}
\item[7.] \pattern{\Q, said Mayor \Sp.}
\item[8--9.] \pattern{\Q, said Mayor of \wildcard \Sp.}
\end{enumerate}
The combination of patterns 3 and 4 allows us to match variations such as \quot{Nauruan writer} and \quot{Martian writer}. Likewise, the clustered patterns 8 and 9 would recognize the Mayor of any city.

In practice, we observe that the wildcards inserted by the algorithm frequently match titles (\eg, \quot{Mrs.}), roles (\eg, \quot{Mayor}, \quot{CEO}), company names, adjectives, and adverbs.
(For additional implementation details, see \Appref{sec:Implementation details}.)

\xhdr{Pattern filtering}
In practice, we typically have thousands of patterns even after the clustering step, and not all of them will be of such high quality as the six patterns from the above example.
Hence, in the final substep of pattern inference, we filter out all patterns having low precision (\Eqnref{eqn:patprec}), \ie, patterns that are not well suited for detecting the speaker of a given quotation.
Pattern filtering is crucial for preventing errors from propagating and becoming amplified in further iterations.
(For additional implementation details, see \Appref{sec:Implementation details}.)

\section{Data}
\label{sec:data}
\subsection{ICWSM 2011 Spinn3r corpus}

We deploy and test the \name algorithm on the ICWSM 2011 Spinn3r dataset \cite{burton2011icwsm}, which consists of news articles, social media posts, forum posts, blog posts, \etc, from the one-month period from January 13 to February 14, 2011.
We limit ourselves to the news portion of the dataset, which comprises about 14 million articles in many languages, but in this evaluation, we focus on English, which is by far the most prevalent language in the corpus.
After removing exact duplicates (\ie, articles having identical contents), we are left with 3.8 million news articles from about 9,500 websites.

The dataset comes with boilerplate elements (headers, footers, sidebars, \etc) removed, though in an imperfect fashion, given the difficulty of reliably detecting such elements for thousands of different websites.
Finally, we turn the remaining portion of each document into plain text by removing HTML tags, scripts, \etc

\subsection{Ground truth of speaker-labeled quotations}
\label{sec:Ground truth}
Our model, being unsupervised, does not require any training set. Nonetheless, we decided to collect an approximate ground truth via crowdsourcing, which is used only for evaluating how well our approach performs.

The most comprehensive way of evaluating the model would be to manually annotate all the quotations, which is obviously unfeasible for such a large dataset, and would defeat the purpose of the algorithm. On the opposite side of the spectrum, the simplest solution would be to evaluate only a small subset of the extracted tuples, which can only tell us about \emph{precision}, \ie, what fraction of the extracted tuples actually represent quotations, and among those, what fraction of speakers are attributed correctly. However, it is equally important to evaluate \emph{recall}, \ie, what fraction of all valid $(Q,S)$ pairs are actually retrieved.

We collected our ground truth in a way that allows us to estimate both metrics. Specifically, we selected a small number of people and constructed a complete dataset containing all of their quotations.
\Tabref{tab:13 people} shows a list of the people in our ground truth. Our motivation for this work was mostly an analysis of political discourse, which is why we put an emphasis on politicians, but we also include four other celebrities to evaluate how well the model generalizes.

Since our algorithm, by design, can handle only direct quotations with explicit mentions, the ground truth comprises only quotations attributed to the speaker via their name (\eg, \quot{\Spcol{John} said}), rather than via anaphora (\eg, \quot{he said}).
In \Secref{sec:Impact of not resolving coreferences}, we discuss how our focus on direct quotations might affect performance, and we argue that it does not represent an issue, as a direct mention is likely to appear at least once in redundant news text.

When constructing our ground truth, we started by extracting a set of \emph{potential quotations}, \ie, all passages in which a quote appears close to a speaker name (within 15 word tokens; \cf Appendix~\ref{sec:Implementation details} for how we detect speakers).
We then had each quotation labeled as either correct or incorrect via crowdsourcing on Amazon Mechanical Turk.\footnote{https://www.mturk.com/} Workers were instructed to label a candidate quotation as correct if it represented a direct quotation \emph{and} the speaker was attributed correctly, and mark it as incorrect if it did not represent a quotation, the quotation was anaphoric, or the speaker was attributed incorrectly. Only ``master workers'' (\ie, experienced annotators) were allowed to participate in the tasks. Each task consisted of 10 candidate quotations, one of which was a test question with an obvious answer. Each task was solved by two independent workers, with an agreement rate of 84\% (Fleiss' kappa 0.63).
In cases of disagreement we broke the tie by collecting a third label.
Most labeling conflicts are due to corner cases that are difficult to annotate, even for humans. An analysis by one of the authors of a small sample of the ground truth (100 quotations) revealed an accuracy of the crowd labeling process of 90\%.

\begin{table}[t]
	\centering
    {\tiny
    \setlength{\tabcolsep}{4pt}
	\begin{tabular}[c]{l|l|r|r|r|r}
		\hline
		\specialcell{Speaker\\} & \specialcell{Profession\\} & \specialcell{Cand.\\quots.} & \specialcell{Actual\\quots.} & \specialcell{Precision\\} & \specialcell{Recall\\}\\
		\hline
		Angela Merkel & Politician & 696 &	269 (38\%) & 98\%  & 30\% \\
		Benjamin Netanyahu & Politician &	618 &	319 (51\%) & 91\%  & 53\% \\ 
		Guido Westerwelle & Politician &	259 & 	178 (68\%) &  98\% & 60\% \\
		John Boehner & Politician &	1,465 &	833 (56\%) & 83\%  & 52\% \\
		John McCain & Politician &	963 &	384 (39\%) & 86\%  & 48\% \\
		M. Ahmadinejad & Politician &	401 &	169 (42\%) & 95\%  & 37\% \\   
		M. ElBaradei & Politician &	1,333 &	583 (43\%) & 99\%  & 37\% \\ 
		Sarah Palin & Politician &	1,919 &	416 (21\%) & 87\%  & 19\% \\
		Vladimir Putin & Politician &	313 &	105 (33\%) & 84\%  & 51\% \\
		Charlie Sheen & Actor &	920 &	102 (11\%) & 100\%   & 15\% \\
		Julian Assange & Org.\ founder &	667 &	153 (22\%) & 97\%   & 29\% \\
		Mark Zuckerberg & Org.\ leader &	362 &	65 (17\%) & 92\%   & 23\% \\
		Roger Federer & Athlete &	344 &	44 (12\%)  & 100\%   & 38\% \\
		\hline
		\specialcell{\\Total} &  & \specialcell{\\10,260} & \specialcell{\\3,620 (35\%)} & \specialcell{Micro 90\%\\Macro 93\%} &	\specialcell{Micro 40\%\\Macro 38\%}\\
		\hline
	\end{tabular}
    }
	\caption{List of speakers in the ground truth, alongside their statistics. Professions were retrieved from the Freebase knowledge database. Precision and recall refer to the evaluation using a single seed pattern (\Secref{sec:Precision and recall on ground truth}).
}
	\label{tab:13 people}
\end{table}

We now focus our discussion on \Tabref{tab:13 people}, which compares the number of candidate quotation with the number of actual (\ie, positively annotated) quotations. The most striking observation is that only 3,620 out of 10,260 samples are correct, which offers an interesting baseline: a simple model that attributes a quotation to the nearest speaker would obtain a precision of only 35\%. We shall compare to this baselines more carefully in \Secref{sec:Precision and recall on ground truth}.

\begin{figure}[t]
	\centering{}
	\includegraphics[width=0.5\linewidth]{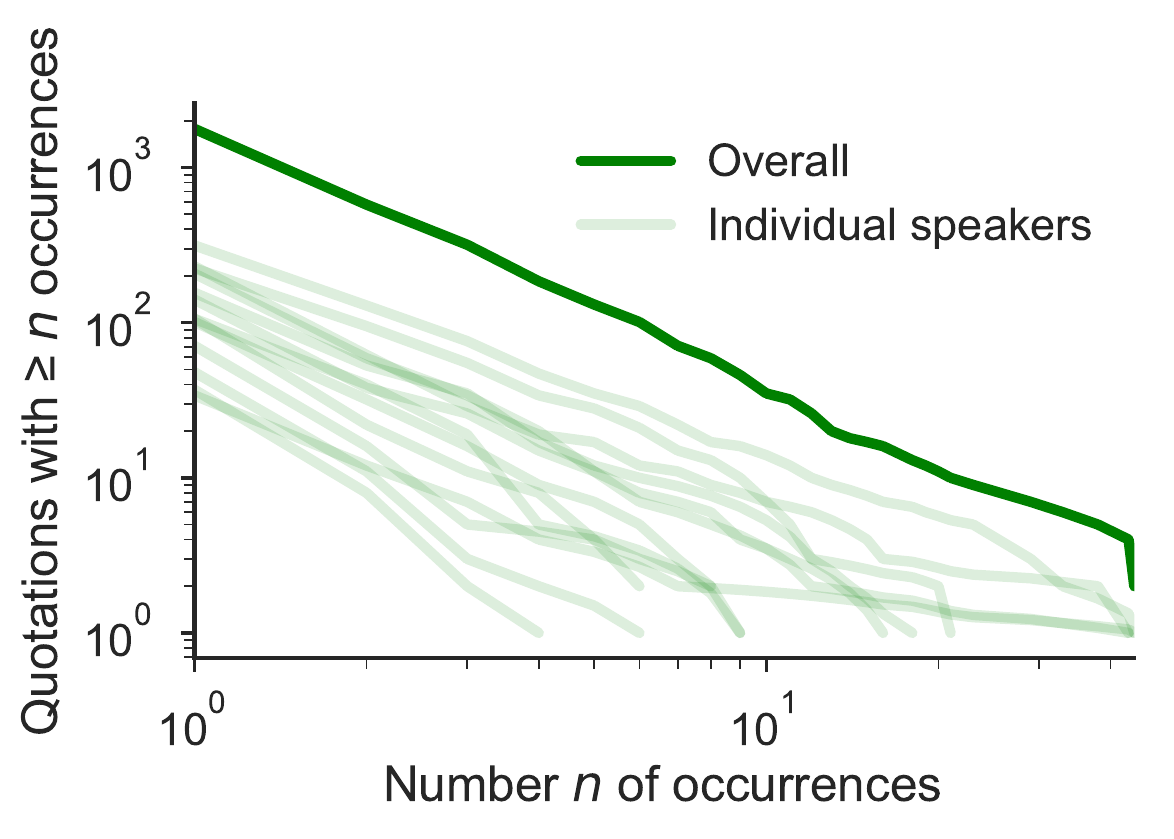}
	\caption{
    Complementary cumulative distribution function (CCDF) (unnormalized) of the number of occurrences per quotation in our ground truth (\Secref{sec:Ground truth}). Light lines represent the distributions of the 13 individual people, while the dark line pools all of their quotations.
    }
	\label{fig:redundancy_ccdf_13people}
\end{figure}

Redundant quotations are counted multiple times in \Tabref{tab:13 people}. In this light, it is worth mentioning that the 3,620 instances stem from just 1,722 unique quotations (see \Appref{sec:Implementation details} for how we group instances). \Figref{fig:redundancy_ccdf_13people} visualizes how the 3,620 instances distribute over the 1,722 unique quotation, via a \emph{complementary cumulative distribution function} (CCDF) for each of our 13 people, as well as for their union. We can observe that most quotations are unique, but there is a long tail of quotations appearing multiple times.
Our algorithm works by boostrapping from these redundant quotations.

\section{Evaluation}
\label{sec:results}
\subsection{Precision and recall on ground truth}
\label{sec:Precision and recall on ground truth}

We use \emph{precision} and \emph{recall} as our main evaluation metrics.
As our aim is not just to detect whether a quotation is retrieved, but also to check if it is attributed to the appropriate speaker, our definitions of precision and recall are slight variations of their usual definitions in an information retrieval context, taking into account both extraction and attribution.

We denote by $\mathcal{X}$ the set of all \emph{unique} $(Q,S)$ pairs in the ground truth (the \emph{relevant quotation--speaker pairs}), which are assumed to have no conflicts, \ie, each quotation is mapped to exactly one speaker, thanks to the conflict resolution step described in \Secref{sec:Conflict resolution}.
We also define $\mathcal{Y}$ to be the set of \emph{retrieved quotation--speaker pairs,} which are also assumed to be unambiguous.
Finally, we introduce $\mathcal{Z}$, the set of retrieved pairs whose $Q$ and $S$ are \emph{both} not in the ground truth. Since we have no annotations for them, they are considered neutral matches.
Our definitions are thus as follows:
\begin{equation}
\operatorname{precision} = \frac{|\mathcal{X} \cap \mathcal{Y}|}{|\mathcal{Y} \setminus \mathcal{Z}|},
\hspace{3mm}
\operatorname{recall} = \frac{|\mathcal{X} \cap \mathcal{Y}|}{|\mathcal{X}|}.
\end{equation}
Intuitively, one pair is considered to be valid if the quotation has been extracted \emph{and} the speaker has been attributed correctly.
Errors may be caused by
\begin{itemize}
\item a relevant quotation that has not been extracted, which causes a decrease in \emph{recall;}
\item a relevant quotation that has been extracted but has been attributed to the wrong speaker (whether relevant or not), which causes a decrease in both \emph{precision} and \emph{recall;} or
\item a non-relevant quotation that has been incorrectly attributed to a relevant speaker, which causes a decrease in \emph{precision}.
\end{itemize}

We compute the precision/recall for each speaker separately, as well as an aggregate precision/recall value for the full ground truth.
With regard to the latter, we provide two versions of our evaluation:
\begin{description}
\item[Micro:] Each $(Q,S)$ pair is weighted equally, thereby taking into account varying quotation counts across speakers.
\item[Macro:] Each speaker is weighted equally, \ie, precision and recall are averaged over all speakers.
\end{description}
The results are summarized in \Tabref{tab:13 people}.
We achieve a precision of 90\% at recall 40\%, even when using only one single seed pattern, \pattern{\Q, \Sp said}, which demonstrates that our algorithm requires minimal human supervision. Another test, using 144 seed patterns including variations of the verb \emph{said} as well as word order permutations, yielded a precision of 91\% at 41\% recall (micro), which is only a slight improvement over the single-seed case.
We conclude that the exact choice of seed patterns does not impact the final result significantly.



\begin{figure*}[t]
    \vspace{-3mm}
    \centering
    \subfigure[]{
		\includegraphics[width=0.23\textwidth]{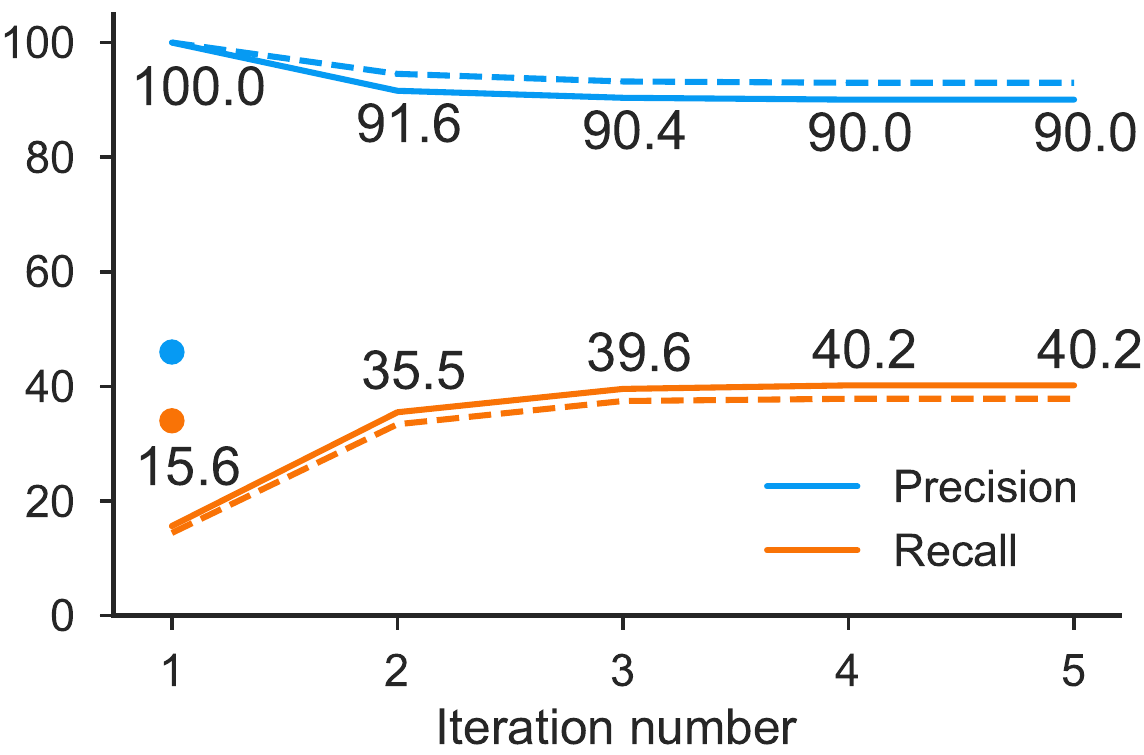}
		\label{fig:prec_rec_13people}
    }
    \subfigure[]{
		\includegraphics[width=0.23\textwidth]{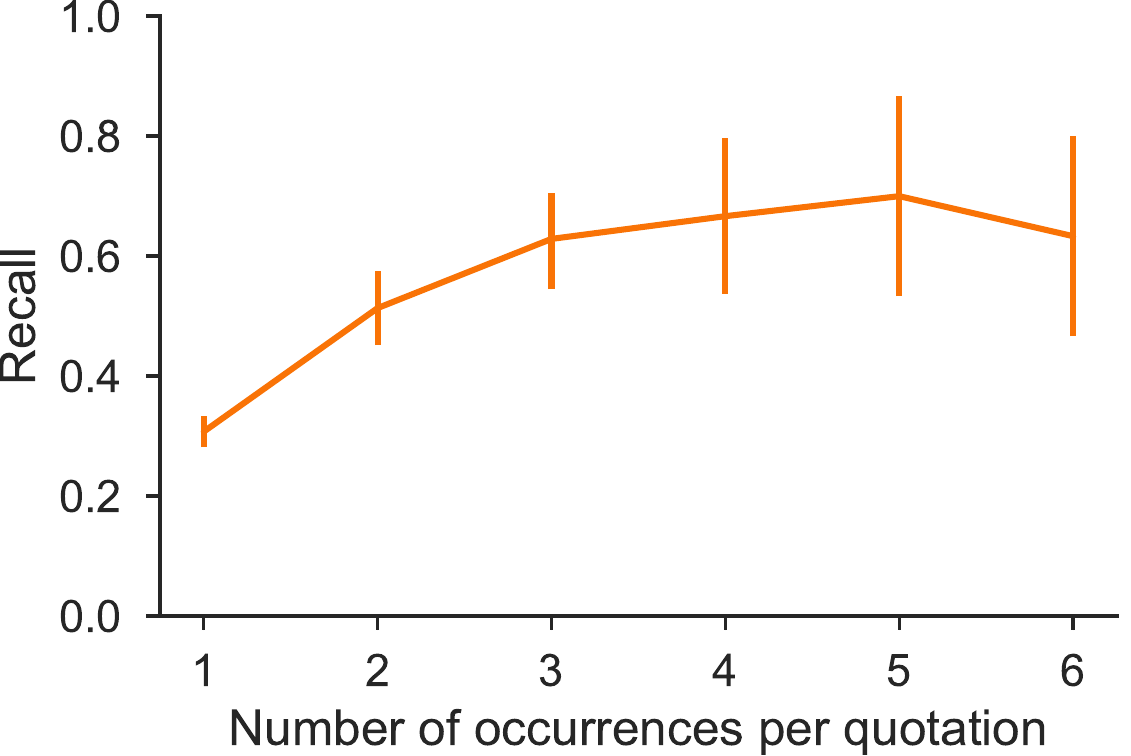}
		\label{fig:rec_by_quot_freq_13people}
    }
    \subfigure[]{
		\includegraphics[width=0.23\textwidth]{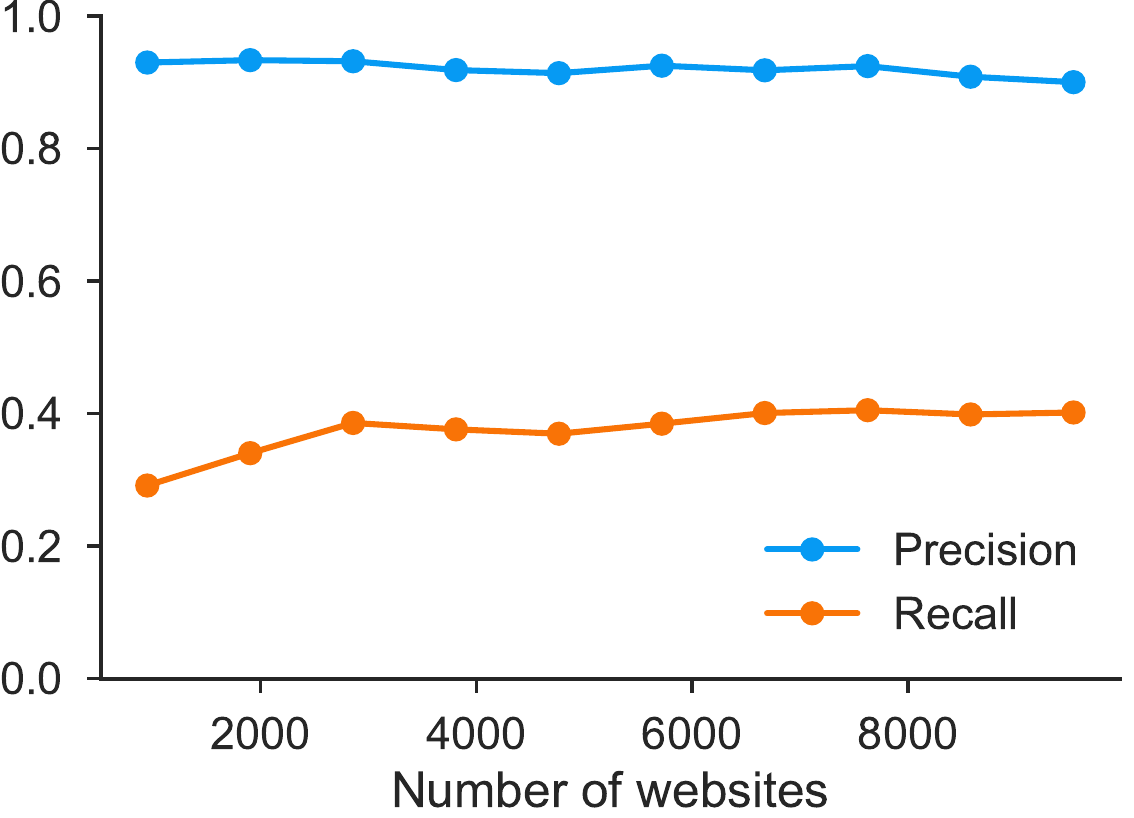}
		\label{fig:prec_rec_by_num_websites_13people}
    }
    \subfigure[]{
		\includegraphics[width=0.23\textwidth]{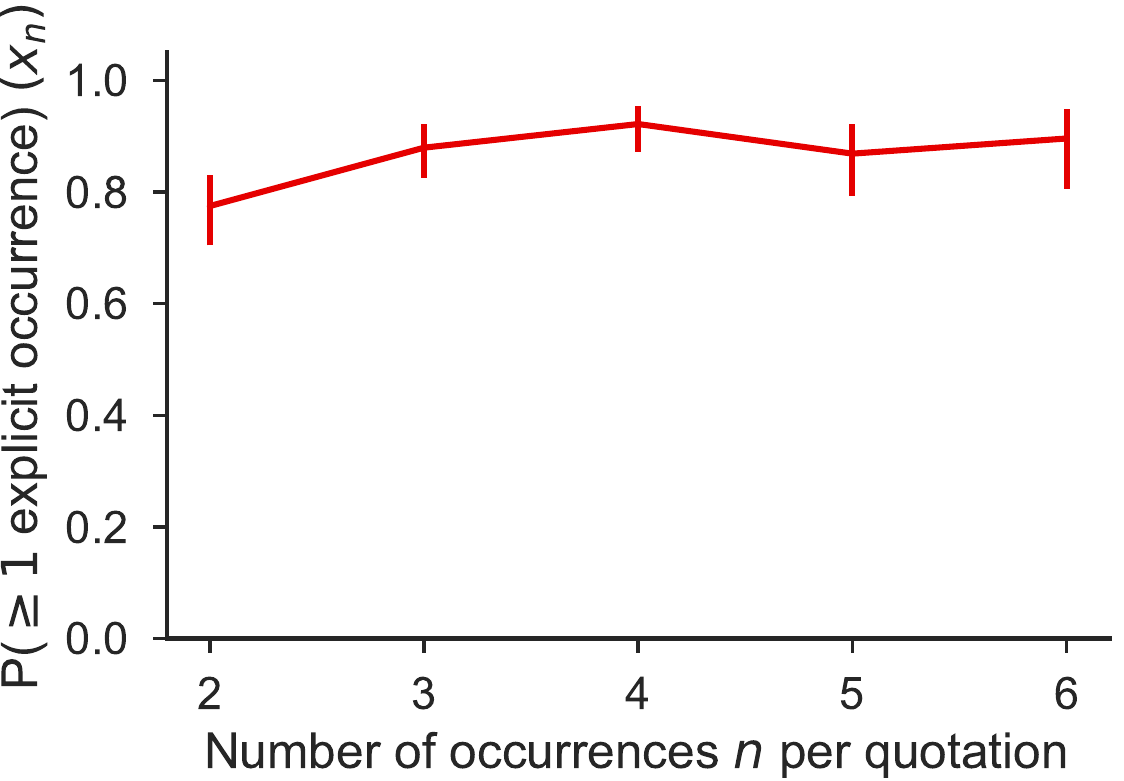}
		\label{fig:quot_prob}
    }
    \vspace{-4mm}
    \caption{
    Results of running \name on our ground truth for 13 speakers (\Secref{sec:Ground truth}).
    \textbf{(a)} Precision and recall (with a single seed pattern) as functions of the number of bootstrap iterations (solid: micro; dashed: macro; dots: nearest\hyp speaker baseline).
    \textbf{(b)} Recall as a function of quotation frequency.
    \textbf{(c)} Precision and recall as functions of the number of randomly sampled news websites from the corpus.
    \textbf{(d)} Probability $x_n$ of observing at least one explicit occurrence for a random distinct quotation with $n$ occurrences (\Secref{sec:Impact of not resolving coreferences}).
    Error bars are 95\% confidence intervals computed via bootstrap (the other bootstrap\dots) resampling.
    }
    \label{fig:results_ground_truth}
    \vspace{-3mm}
\end{figure*}

Since our algorithm is iterative (it consists of multiple cycles of the loop depicted in \Figref{fig:pipeline}), it is natural to ask how precision/recall evolves as we perform more cycles. \Figref{fig:prec_rec_13people} plots these metrics as a function of the number of iterations. We can observe that the bootstrapping process is fast and converges by the fourth iteration. Recall increases sharply, whereas precision decreases much less drastically, showing the effectiveness of bootstrapping, especially when combined with thorough filtering (\Secref{sec:Pattern inference}).

It is also interesting to assess how our method relates to the baseline that we used for extracting the ground truth in \Secref{sec:Ground truth}. To this end, we replaced the pattern matching step with a heuristic that attributes a quotation to the nearest speaker, and we kept the conflict resolution step. This baseline yields a precision of 46\% at 34\% recall, which is not acceptable for practical purposes. Note that the 35\% precision figure reported in \Secref{sec:Ground truth}, despite being related, is fundamentally different from that of this evaluation. It is important to distinguish between per-instance evaluation and per-quotation evaluation (which is what we are performing). A per-instance evaluation of precision/recall in the context of news articles does not take into account the ultimate goal of a quotes dataset (attributing a quotation to exactly one speaker), and does not define a precise behavior in case of conflicts. With our definition, an unresolved conflict causes a decrease of both precision and recall. For the same reason, the baseline does not reach a recall of 100\%: a quotation is labeled as valid only if it is attributed to the proper speaker.

\subsection{Impact of not resolving coreferences}
\label{sec:Impact of not resolving coreferences}
An evident downside of our algorithm is that the learned patterns are geared toward \textit{explicit speaker mentions} (\eg, \quot{\Qcol{``I kill-e''}, said \Spcol{Queequeg}.}), and cannot immediately handle anaphora (\eg, \quot{\Spcol{Queequeg} startled. \Qcol{``I kill-e''}, he said.})
One way of addressing this is to run a coreference resolver as a preprocessing step, but unfortunately, coreference resolution algorithms perform poorly, especially on languages other than English \cite{pouliquen2007automatic}.

Instead, we argue that the lack of coreference resolution does not hurt our model significantly: given the redundant nature of large news corpora, many quotations are reported multiple times, and if at least one of the occurrences is explicit (\ie, mentions the speaker by her name, rather than a pronoun), our algorithm can potentially detect it.

To make this argument more formal, we would like to know what fraction $x_n$ of all distinct quotations with $n$ total occurrences has at least one explicit occurrence (and can thus in principle be discovered by \name).
Unfortunately, we cannot immediately derive the denominator of the fraction $x_n$ from the ground truth, as we do not know the number of distinct quotations with \textit{no} explicit speaker mention (since the ground truth contains only explicit quotation occurrences).
One quantity we can derive from the ground truth, however, is the probability $p_n$ that, if we randomly sample one of a quotation $Q$'s $n$ occurrences, the sampled occurrence is explicit, \textit{given that $Q$ has at least one explicit occurrence} (since only then is it contained in the ground truth).
And luckily, we can estimate $x_n$ from $p_n$ and $n$, as follows.
First, note that, if we sample a random occurrence from across all quotations with exactly $n$ occurrences, the probability that the sampled occurrence is explicit is given by the product $p_n x_n$.
Expressing the probability that a quotation has at least one explicit occurrence (\ie, $x_n$) in terms of the probability that a randomly drawn occurrence is explicit (\ie, $p_n x_n$), we have
\begin{eqnarray}
x_n &=& 1 - (1-p_n x_n)^n.
\end{eqnarray}
For $n>1$, this polynomial equation has exactly one solution $x_n$ in the interval $]0,1]$.
\Figref{fig:quot_prob} plots our estimate of $x_n$ as a function of $n$. As expected, $x_n$ increases with $n$, and, most important, it is high for all $n\geq 2$, namely 77\% for $n=2$, and over 80\% for all $n>2$.

We conclude that not doing coreference resolution does not represent a major problem in our scenario, as the redundancy in the dataset makes explicit mentions likely to appear.

\begin{figure}[t]
    \vspace{-3mm}
    \centering
    \subfigure[Quotations per person]{
        \includegraphics[width=0.22\textwidth]{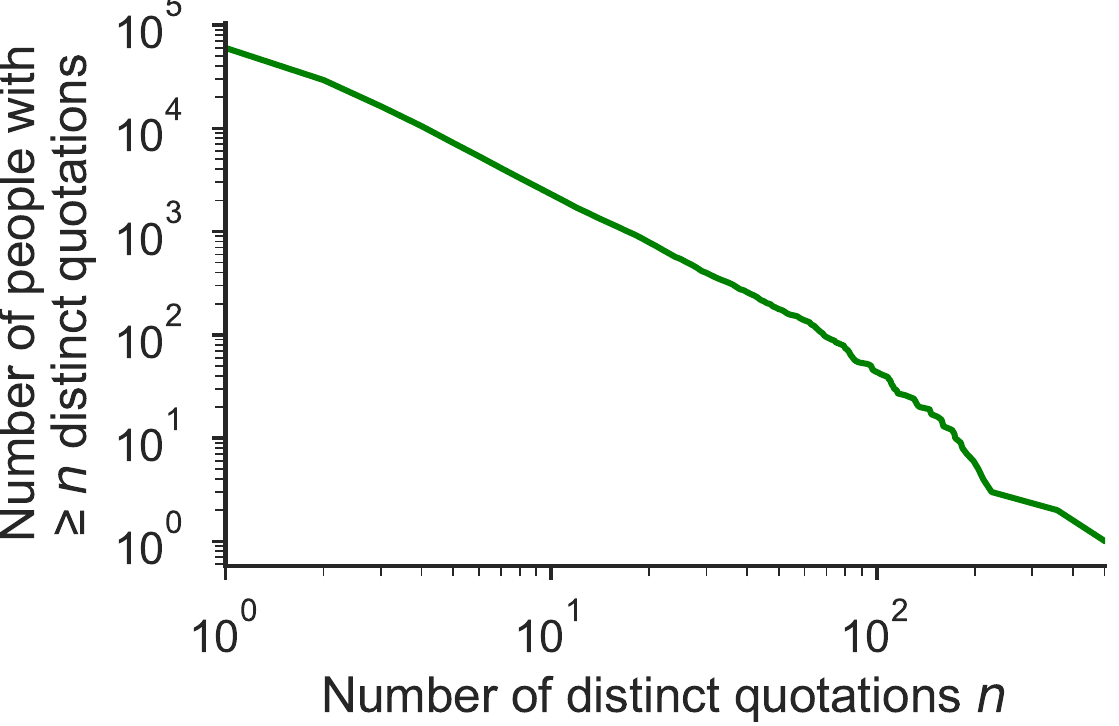}
        \label{fig:people_quotations_ccdf}
    }
    \subfigure[Quotation redundancy]{
        \includegraphics[width=0.22\textwidth]{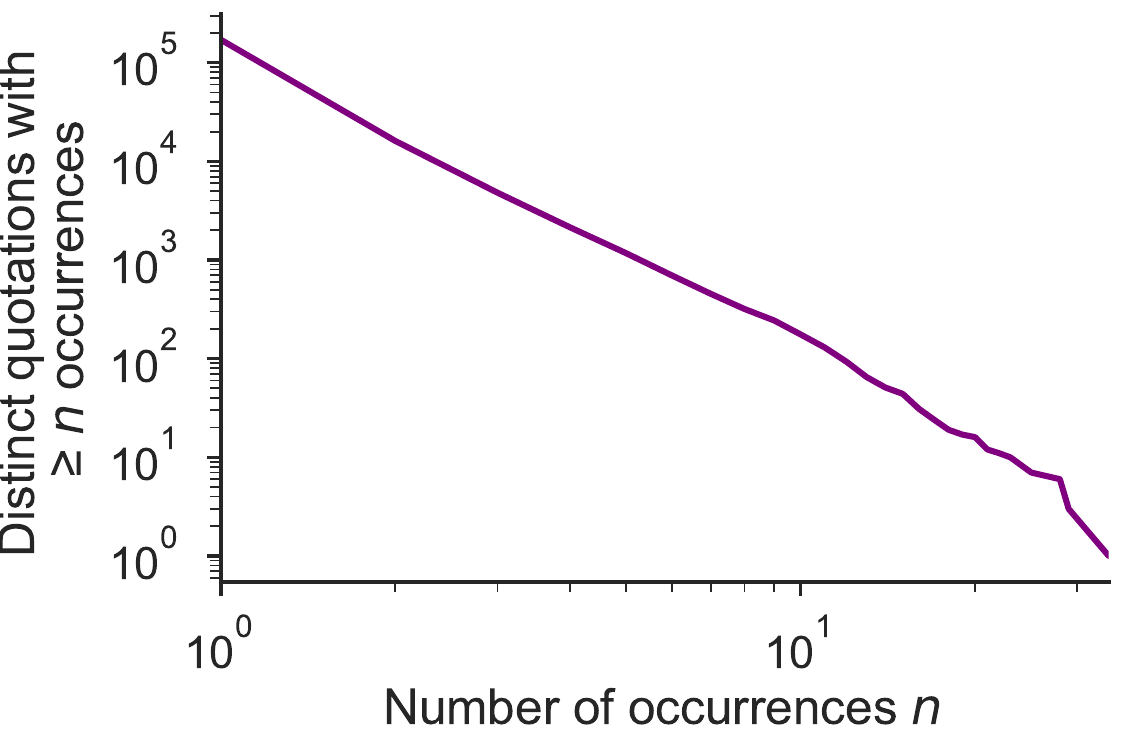}
        \label{fig:quotations_redundancy_ccdf}
    }
    \vspace{-4mm}
    \caption{
    Complementary cumulative distributions computed on the $(Q, S)$ pairs extracted from the full corpus.}
    \vspace{-3mm}
\end{figure}

\subsection{Quotation from full ICWSM 2011 dataset}
\label{sec:Full ICWSM 2011 quotation dataset}



The evaluation in \Secref{sec:Precision and recall on ground truth} confirms that the method works well on a small set of 13 hand-picked people. 
Here, we generalize the analysis by considering all \qsp{}s extracted from the ICWSM 2011 Spinn3r corpus.

\xhdr{Basic statistics of quotation dataset}
The resulting dataset consists of 171K quotations, attributed to 60K different people (\Appref{sec:Implementation details} explains how we detect names and link them to the Freebase knowledge base).

The most substantial portion of the dataset comprises people with only few quotations reported by the press in the time-span considered. These people represent a long tail of globally insignificant speakers who appear in the media only sporadically, or are associated with local news. \Figref{fig:people_quotations_ccdf} summarizes the number of quotations per person as a CCDF, and reveals that 51\% of the people found in $(Q,S)$ pairs extracted by our algorithm have exactly one quotation.

A similar pattern is observable in \Figref{fig:quotations_redundancy_ccdf}, which shows the CCDF of the number of occurrences per quotation. A substantial portion (91\%) of the extracted quotations was found in only one instance across all websites. Since \name leverages the redundancy of the quotations to discover new patterns via bootstrapping, it is worth noting that the remaining 9\% of the element represents a critical portion for the algorithm.

\xhdr{Impact of dataset size}
\Figref{fig:prec_rec_by_num_websites_13people} summarizes the impact of the dataset size on the precision and recall computed on our ground truth. To simulate the scenario of a smaller dataset, we subsampled the websites represented in the Spinn3r dataset in increments of 10\%. This evaluation provides an estimate of the number of websites necessary for the bootstrapping mechanism to work. The plot shows that \name can keep its performance intact by using just 30\% (about 3K) of the English websites in Spinn3r. This result provides an insight into the generalizability of our method to other languages, which have lower numbers of online newspapers than English.


\xhdr{Speaker statistics}
The corpus shows a high gender imbalance between men (66\%) and women (34\%), in line with findings of prior work on imbalance in knowledge bases \cite{DBLP:journals/corr/WagnerGJS15}.
Also, since we focus our analysis on English articles, there is a news coverage bias toward the United States. The speaker with most reported quotes is Barack Obama (502), followed by his U.S. press secretary Robert Gibbs (358) and Hillary Clinton (225). Interestingly, politicians are followed by sports coaches, led by Mike Tomlin (212) and Bill Self (205). This could be related to Super Bowl XLV, which happened during the one-month period spanned by our corpus.
The impact of this event is also reflected when aggregating by profession using the information from Freebase: the most represented profession is athletes (23\% of people with at least one quotation), followed by authors (16\%) and politicians (12\%);
24\% of people in our output have no known profession in Freebase.

The most reported quotation is a sentence of then--Russian president Dmitry Medvedev about the January 24, 2011, terrorist attack at Moscow airport (``From the preliminary information we have, it was a terror attack.'') with coverage on 34 different news websites.

\xhdr{Precision}
The evaluation methodology of \Secref{sec:Precision and recall on ground truth} was based on a set of 13 prominent speakers, \ie, speakers with many quotations. To see how well our model performs with respect to all speakers, we performed the simplest evaluation mentioned in \Secref{sec:Ground truth}:
we selected a sample of the extracted \qsp{}s and evaluated their precision using human annotators. Similarly to our ground-truth collection, we performed this evaluation on Amazon Mechanical Turk.

In order to have a fair comparison, we ran this evaluation using both the baseline method (nearest speaker) and our method. We stratified speakers in 9 buckets by quotation count ($2^i$ for $i=1,\dots,9$), with the goal of covering all speaker types in the population. Then we sampled 10 people within each bucket and selected all of their quotations, up to a limit of 50 quotations per person. The resulting 2,400 tuples were annotated by 5 workers each, which produced a Fleiss' kappa of 0.59, slightly lower than the ground truth annotation (maybe due to the fact that we did not request ``master workers'' for this task).

\begin{figure}[t]
	\centering{}
	\includegraphics[width=0.44\linewidth]{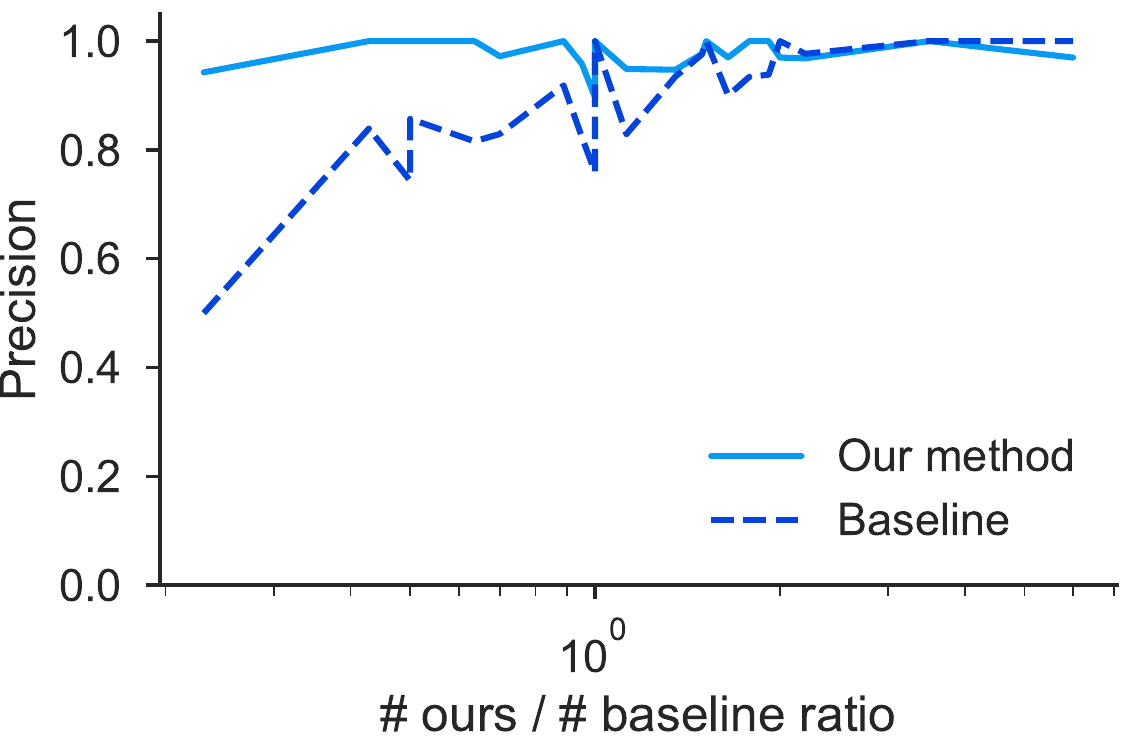}
	\caption{
    Precision of nearest-speaker baseline and of our method, over a set of speakers stratified by the imbalance in the number of quotations extracted by our \vs the baseline method (\cf\ \Secref{sec:Full ICWSM 2011 quotation dataset}). The baseline is confused when other speakers are mentioned in the context.
    }
	\label{fig:baseline}
\end{figure}

Surprisingly, we observed a precision of 80\% for the baseline and 98\% for our method, which is significantly different from our prior evaluation. In an attempt to investigate this, we noticed that famous speakers tend to be mentioned frequently close to other people, whereas less known speakers (\ie, those with few quotations) tend to be unique in a given context.
Hence, in \Figref{fig:baseline}, we stratify speakers by the ratio of the number of quotations extracted by our method over the number of quotations extracted by the baseline method, and plot precision as a function of this ratio.
It is easy to see that a low ratio indicates that a speaker appears often near other people (who actually uttered the respective quotations), which in turn confuses the baseline, such that the latter cannot trade off precision for recall accurately.
Our method, on the other hand, maintains consistent performance. Among the speakers in the lower buckets, we can identify people such as Barack Obama, Hosni Mubarak, Ronald Reagan, and even speakers from our ground truth (\eg, Sarah Palin and Mohamed ElBaradei), which explains why the baseline performs poorly in our ground-truth evaluation (\Secref{sec:Precision and recall on ground truth}).

\section{Example application: sentiment analysis}
\label{sec:apps}


\begin{table}[t]
    
    \begin{minipage}{.5\linewidth}
      
      \centering
    {\tiny
    \setlength{\tabcolsep}{4pt}
        	\begin{tabular}[c]{l|r|r|r}
		\hline
		Profession & \specialcell{Mean\\sent.} & \specialcell{Std.\\dev.} & \specialcell{Num.\\people} \\
		\hline
Politician & 0.175 & 0.26 & 1,837\\
Author & 0.21 & 0.31 & 1,003\\
Actor & 0.26 & 0.29 & 677\\
Artist & 0.27 & 0.30 & 484\\
Athlete & 0.30 & 0.24 & 2,359\\
Athl.\ coach & 0.31 & 0.18 & 448\\
Org. leader & 0.43 & 0.27 & 784\\
		\hline
	\end{tabular}
    }
    \end{minipage}%
    \begin{minipage}{.5\linewidth}
      \centering
        
        \centering
    {\tiny
    \setlength{\tabcolsep}{4pt}
        		\begin{tabular}[c]{l|r|r|r}
		\hline
		Name & Sent. & \specialcell{Std.\\dev.} & \specialcell{Num.\\quot.} \\
		\hline
    Denny Doyle & 0.63 & 0.17 & 21\\
    David Poile & 0.50 & 0.22 & 22\\
    Steven Ballmer & 0.45 & 0.25 & 25\\
    \hline
    Vladimir Putin & $-0.05$ & 0.63 & 35\\
    Jan Brewer & $-0.02$ & 0.64 & 27\\
    Ehud Barak & 0.27 & 0.66 & 25\\
		\hline
	\end{tabular}
    }
    \end{minipage} 
    \caption{\textbf{Left:} Sentiment scores for the most frequent professions (based on people with at least three quotations). \textbf{Right:} Sentiment scores of top three people with lowest and highest standard deviation of sentiment scores.}
    \label{tab:sentiment}
\end{table}

The quotation database extracted by \name enables a vast number of possible use cases. The purpose of this paper is primarily to introduce the \name algorithm, and not to analyze the resulting dataset.
Nonetheless, in this section we briefly scratch the surface of what can be done with a simple analysis of the sentiment conveyed by quotations reported by the press.
We use \textit{VADER} \cite{ICWSM148109} to extract the sentiment score of all retrieved quotations. We adopt the \emph{compound value,} a convenient normalized score between $-1$ (extremely negative) to 1 (extremely positive) that describes the polarity of sentences. We remove quotations with a neutral sentiment of 0 (25\% of all quotations) from the original dataset because, generally speaking, they are too short in terms of text to extract meaningful keywords and compute the score. The neutral quotations have an average length of 79 characters, as opposed to 123 for polarized statements.
By exploiting the mapping with Freebase, we can access additional information about the speaker, such as profession, gender, and age.
\Tabref{tab:sentiment} (left) shows the average scores for the most frequent professions based on the people with at least three quotations.
In general, the sentiment score of the quotations reported in the news tends to be positive. According to these results, politicians tend to make more neutral statements. On the other hand, \emph{organizational leaders}, which include companies CEOs, tend to speak very positively, with a 2.4-fold increase over politicians' scores.

To investigate the average sentiment of popular speakers, we selected people with a significant number (over 20) of quotations in the analyzed month. The person with the lowest sentiment score is Eric Holder ($-0.36$), who served as the United States Attorney General at the end of January 2011 and made many statements about crime and terrorism. On the opposite side of the spectrum, there is another politician, Lisa P. Jackson (0.69), who was head of the Environmental Protection Agency (EPA) at that time.

Additionally, we can detect the people with the most stable attitude across different quotations by measuring the standard deviation of their sentiment scores. \Tabref{tab:sentiment} (right) summarizes the top three most consistent and inconsistent people in terms of the words used in the analyzed month.
The most consistent people are baseball player Denny Doyle, followed by hockey team manager David Poile and Microsoft CEO Steven Ballmer. 
Interestingly, the people with the most volatile language are all politicians: Ehud Barak (Israeli Minister of Defense), Jan Brewer (Governor of Arizona), and Vladimir Putin (Prime Minister of Russia in 2011).

A comparison between males and females does not produce a statistically significant difference. On the other hand, by stratifying the people by age, we can observe a more positive language for young speakers, although this is correlated with the fact that they are more likely to be athletes.




\section{Discussion and future work}
\label{sec:discussion}
We presented \name, an unsupervised method for extracting \qsp{}s from large news corpora. Our approach requires minimal human assistance and reaches 90\% precision at 40\% recall on the ICWSM 2011 Spinn3r dataset by using a single seed pattern. We proposed a novel method for inferring new patterns, as well as an evaluation methodology suitable for large, redundant datasets. Finally, we demonstrated the usefulness of our work by proposing and exploring some potential data analysis tasks.

Compared to a machine\hyp learning--based approach, ours presents several advantages:
\begin{itemize}
	\item \textit{Language independence:} \name was designed to be able to adapt to multiple languages, provided that the seed patterns are chosen accordingly.
	\item \textit{White-box model:} The learned patterns are clearly interpretable and easy to evaluate.
	\item \textit{Reusability:} The learned patterns could be exported and used for completely different applications or in other datasets, as regular\hyp expression matchers are widespread.
	\item \textit{Unsupervised learning:} Except for the few seed patterns, minimal human intervention is required. \name does not require annotating a training set by hand.
\end{itemize}
The major theoretical drawback of \name is its inability of resolving coreferences, but we showed that this is not a major issue in practice. Nonetheless, coreference resolution can be added as a preprocessing step of our pipeline.

Future work on this subject should certainly focus on other languages, especially pronoun\hyp dropping ones, where our model is expected to excel over methods that do not exploit quotation redundancy. To analyze the evolution of political speech over time, it would also help to use a dataset that spans multiple years, instead of one month.

A final interesting direction would be to explore the use of the quotation dataset extracted by \name for training supervised machine learning models, in a process known as \emph{distant supervision} \cite{mintz2009distant}.

\xhdr{Efficient distributed implementation in Spark}
Due to the size of typical news corpora (\eg, the ICWSM 2011 dataset we used here comprises 3.1 TB), it is important to have an efficient and scalable implementation of our algorithm.
We developed a software package in Java for the distributed Spark platform.
It implements an efficient indexing system to map quotations to the contexts in which they appear, and handles patterns matching, the most costly operation in our loop, by leveraging efficient data structures such as hash tries.
We release all our code and data at
\githubrepo.

\bibliographystyle{aaai}
\bibliography{quootstrap}

\appendix
\section{Implementation details}
\label{sec:implementation}
\label{sec:Implementation details}

\xhdr{Detection of speaker names}
A common approach is to use \emph{named entity recognition} (NER).
With language independence in mind, we, however, opted for an approach that does not rely on NER, as NER would need to be trained separately for each language. Instead, we use Freebase, a knowledge base that contains names and aliases (\eg, John Sidney McCain III, John McCain) for about 3 million people and links all names of the same person to a unique ID.

Our implementation also takes care of partial mentions of a name (\eg, McCain \vs John McCain) by finding unambiguous \emph{superstrings} (\ie, the full name of the person) in the rest of the article. If we find such a match, we expand the partial name to the full name. If there are multiple possible matches, the tuple is simply discarded.

\xhdr{Grouping versions of same quotation}
Despite the dataset being redundant, many quotations are not exact duplicates; in some cases, they are phrased slightly differently, and in other cases they are split into multiple smaller quotations.
We address both cases using a single strategy, namely by grouping quotations that share at least one substring of $\ell$ tokens (we use $\ell=8$), and by substituting them with the longest quotation within the group.
The result on the Spinn3r dataset is that 20\% of quotations are merged. We found the impact of grouping \vs not grouping on the final evaluation to be small (about 1\% improvement in precision and recall), but it might make a difference for individual quotations, so we decided to include this step in our pipeline.

\xhdr{Pattern clustering}
This step increases the generality of patterns by replacing individual tokens with wildcards \wildcard if their occurrence count is below a threshold $n_{\min}$ (\Secref{sec:Pattern inference}).
In practice, we use relative (rather than absolute) thresholds, with respect to the total number of unclustered patterns $N$. Furthermore, we use multiple thresholds ($n_{\min} = 0, 0.0002 \, N, 0.001 \,N, 0.005 \,N$), whose resulting patterns are combined together.
Different thresholds trade off precision for recall in different ways (a lower threshold means lower precision but higher recall). Note, however, that including lower thresholds does not impact the algorithm negatively, as it is protected by the pattern filtering step (\Secref{sec:Pattern inference}).

\xhdr{Pattern filtering}
Our definition of pattern precision (\Eqnref{eqn:patprec}) is solely a function of the numbers of positive and negative matches.
In practice, we also weight each match according to the length of the corresponding quotation (with higher weights for longer quotations), since short quotations are more likely to have been uttered by several people, whereas long quotations tend to be unique to a speaker.
We use a $\tanh$ function on the length of the quotation, such that quotations of length 0 get weight 0, and the weight of longer quotations tends to 1 as their length approaches infinity.

Finally, for the choice of the filtering threshold, we find values between 0.7 and 0.95 to give good results, depending on the desired trade-off between precision and recall. In all our experiments, we used a fixed threshold of 0.7. We also discard patterns that extract fewer than $M$ previously discovered \qsp{}s, as they produce unreliable estimates of pattern precision (we use $M=5$).







\end{document}